\documentclass[twocolumn]{aastex631}

\usepackage{ulem}
\usepackage{amsmath}

\defcitealias{2015AJ....150..189G}{G15}
\defcitealias{2014ApJ...780..172D}{D14}

\shorttitle{A Test for Systematics in the JAGB Method}
\graphicspath{{./}{figures/}}

\begin{document}

\title{Carbon Stars as Standard Candles:\\ An Empirical Test for the Reddening, Metallicity, and Age Sensitivity of the \\J-region Asymptotic Giant Branch (JAGB) Method} 

\author{Abigail~J.~Lee}\affil{Department of Astronomy \& Astrophysics, University of Chicago, 5640 South Ellis Avenue, Chicago, IL 60637}\affiliation{Kavli Institute for Cosmological Physics, University of Chicago,  5640 South Ellis Avenue, Chicago, IL 60637}

\correspondingauthor{Abigail J. Lee}\email{abbyl@uchicago.edu}

\begin{abstract}
The J-region Asymptotic Giant Branch (JAGB) method is a standard candle based on the  intrinsic luminosities of carbon stars in the near infrared. 
We directly constrain the impact of metallicity, age, and reddening on the JAGB method.
We assess how the mode, skew, and spread of the JAGB star luminosity function change throughout diverse stellar environments in M31's NE disk from $13<d<18$ kpc using data from the \textit{Panchromatic Hubble Andromeda Treasury} (PHAT). 
As expected, the mode is found to be fainter in higher-reddening regions. 
To cross-check this result, we also measure a fiducial J-band ground-based JAGB distance using data from the UKIRT/WFCam in M31's outermost disk ($18<d<40$~kpc) where internal reddening is minimal. 
We find that this J-band distance modulus agrees well with the F110W distance moduli measured in the lowest reddening regions of the PHAT data, demonstrating the JAGB method is most accurate if measured in the low-reddening outer disks of galaxies.
On the other hand, the mode of the JAGB star luminosity function appears empirically to show no dependence on age or metallicity within the range $-0.18<[M/H]<-0.26$~dex.
In conclusion, the JAGB method proves to be a robust standard candle capable of calibrating the luminosities of type Ia supernovae and therefore providing a high-accuracy, high-precision measurement of the Hubble constant. 
\end{abstract}


\section{Introduction}

The main source of the 2$\sigma$ difference between the Cepheid and tip of the red giant branch (TRGB) measurements of $H_0$ lies in their disagreeing calibrations of the SNe Ia at distances in excess of 10 Mpc. This indicates attention must be focused on scrutinizing the measurements of these distances for undetected systematic errors from both the Cepheids and the TRGB. 
However, no other distance indicator has been both bright enough to measure distances farther than 10 Mpc and capable of being measured precisely enough to act as a third cross check. That is, until the J-region Asymptotic Giant Branch (JAGB) method was recently introduced and has repeatedly proven its accuracy as a distance indicator multiple times, by several independent groups \citep{2020ApJ...899...67F, 2020MNRAS.495.2858R, 2021arXiv210502120Z}. 

The JAGB method measures distances via a subset of carbon-rich, thermally-pulsating AGB (TP-AGB) stars, or \textit{carbon stars}. These carbon stars have low-dispersion absolute magnitudes in the near infrared, making them excellent standard candles. 
The JAGB method has been shown to be competitive in measuring distances with the Cepheid and TRGB distance indicators in both precision ($<4$\% uncertainty on individual distances; \citealt{2021ApJ...907..112L, lee2022, 2021MNRAS.501..933P, 2023MNRAS.522..195P}) and accuracy (distances measured via the JAGB method agree with independent TRGB and Leavitt law distances to within a combined scatter of $\pm0.08$~mag; \citealt{2020ApJ...899...67F, 2021arXiv210502120Z}). 
JAGB stars are thus sufficiently accurate enough to act as a cross-check on SN Ia host galaxy distances measured by the TRGB and Cepheids, particularly for measuring the Hubble constant ($H_0$). 

The narrow range of carbon star luminosities has been predicted by stellar evolution theory for nearly half a century \citep{1973ApJ...185..209I}. 
TP-AGB stars transition from oxygen-rich (where the ratio of carbon-to-oxygen in their atmospheres is C/O$<1$) to carbon-rich stars (C/O$>$1)  following the \textit{third dredge-up event}, during which  thermal pulsations dredge  up carbon from the helium-burning shell out onto the surfaces of these stars. The stars' resulting carbon absorption bands, mainly CN and $\rm{C_2}$, reside in the middle of traditional photometric bandpasses and therefore strongly affect the carbon stars' color, making them demonstrably redder than typical oxygen-rich AGB stars. Furthermore, the third dredge-up event is only effective for AGB stars with a narrow range of masses ($2-5 M_{\odot}$), and thus, luminosities. 
Younger, more massive stars undergo hot bottom burning (HBB) where the carbon is converted to nitrogen before it reaches the surface, and the dredge-up for older, less massive stars does not operate efficiently enough due to the progressive thinning of the stellar envelope \citep{2008A&A...482..883M}. 
These theoretical limits are observationally reflected in the well-defined magnitude and color bounds of carbon stars, which can be further segregated into a subclass of \textit{JAGB stars} through simple color cuts which distinguish them from O-rich AGB stars (to the blue) and extreme AGB stars (to the red), respectively. 
The peak location of the JAGB star luminosity function then marks the apparent JAGB magnitude. Identifying JAGB stars requires only two-band, single-epoch photometry, making this method of measuring distances extremely simple and observationally efficient. 

Because the JAGB method is relatively novel, it still needs to be scrutinized carefully for systematics. For example, a host galaxy's age gradient, metallicity, and internal reddening could affect the shape and peak location of the JAGB star luminosity function, potentially introducing unwanted systematic errors in measured distances. 

Theory has predicted the metallicity of the galaxy could have an effect on the magnitudes of the carbon stars. Using synthetic AGB star models, \cite{1999A&A...344..123M} first attributed differences in the location of the peaks in the CSLFs of the LMC vs. the SMC to their different metallicities. 
These prediction was supported again recently by \cite{2020MNRAS.498.3283P}, who used the the \textsc{trilegal} stellar population synthesis simulations to study the shapes of CSLFs in the Magellanic Clouds.
In both studies, the peak location of the CSLF in the more metal-rich LMC was found to be brighter than the peak location of the more metal-poor SMC's CSLF, due to the dependence of the third dredge-up's efficiency on the metallicity of its stellar environment. 
As metallicity increases, the dredge-up efficiency decreases, and therefore the minimum mass for a star to experience dredge-up of carbon increases \citep{2003agbs.conf.....H}. 
For example, at an initial metallicity of $Z_i=0.004$, carbon stars formed at initial masses of $M_i\approx1.4M_{\odot}$ to  $M_i\approx2.8M_{\odot}$. For an initial metallicity of $Z_i=0.008$, carbon stars formed at more massive initial masses of $M_i\approx1.7M_{\odot}$ to  $M_i\approx3M_{\odot}$ \citep{2020MNRAS.498.3283P}. Since the initial mass of a star translates into its luminosity, the more metal-rich carbon stars would be expected to have a brighter average luminosity.

Recent papers by the group of \cite{2020MNRAS.495.2858R, 2021MNRAS.501..933P, 2023MNRAS.522..195P} found the more metal-rich LMC's carbon star LF had a mode of $M_J=-6.33\pm0.01$~mag, whereas the more metal-poor SMC's carbon star LF had a mode of $M_J=-6.18\pm0.01$~mag. These empirical measurements agreed with the theoretical predictions of \cite{1999A&A...344..123M,2020MNRAS.498.3283P}. However, other studies in the literature instead have found excellent agreement between JAGB measurements in the different Magellanic Clouds. Using the mean, \cite{2020ApJ...899...66M} measured  $M_J=-6.22\pm0.03$~mag for the LMC and $M_J=-6.18\pm0.05$~mag for the SMC. Similarly, using the mean to measure the JAGB magnitude, \cite{2021arXiv210502120Z} measured $M_J=-6.21\pm0.03$~mag for the LMC, and $M_J=-6.20\pm0.04$~mag for the SMC.

Indirect tests indicate metallicity is likely not a significant source of systematic error ($<0.06$~mag) on the measurement of distances via the JAGB method.  \cite{2022ApJ...926..153M} compared distances to 34 galaxies from the JAGB method vs. TRGB, finding an inter-method scatter of only $\pm0.08$~mag. Assuming the scatter is shared equally between the two methods, they would each have a precision of $\pm0.06$~mag, meaning any effects from the host galaxy star formation history and/or metallicity differences are constrained at this level of significance. Similarly, \cite{2021arXiv210502120Z} compared distances to eight galaxies using the Cepheid period-luminosity relation and JAGB method, finding an inter-method scatter of $\pm0.09$~mag, delivering a similar individual precision of $\pm0.06$~mag for each distance indicator. 
An additional useful test for unearthing systematics in the JAGB method is to compare distances from all three methods applied to the same galaxy. \cite{2021ApJ...907..112L, lee2022} measured TRGB, JAGB, and Cepheid P-L distances to WLM and M33, respectively, using high-precision observations specifically optimized for each distance indicator. They found agreement at the 3\% and 2\% level respectively, indicating any astrophysical systematics in the JAGB method must be small, given that the JAGB stars, red giant branch stars, and Cepheids were drawn from entirely independent stellar populations and are thus expected to be influenced differently by metallicity.

In this paper, we directly empirically test for these second-order effects using the wealth of near-infrared and optical \textit{Hubble Space Telescope} (HST) data from the \textit{Panchromatic Hubble Andromeda Treasury} (PHAT).

More than three decades ago, \cite{1990ApJ...365..186F} published the first empirical test for a metallicity effect on the Cepheid period-luminosity relation in the Andromeda galaxy (M31). M31 is an ideal galaxy for testing distance indicators because it is nearby, hosts diverse stellar populations, and suffers from low Galactic foreground extinction. M31's distance is also uncontroversial and well-agreed upon, allowing us to compare our measured distances to many others in the literature.
In this paper, we return to  Andromeda to study the astrophysical systematics of the JAGB method. 
Rigorously testing the JAGB method with HST will then set the stage for even farther JAGB distances to be measured with the James Webb Space Telescope and Nancy Grace Roman Space Telescope, both optimized for the NIR wavelengths where carbon stars are their brightest.

\section{Data}\label{sec:phot}

The PHAT survey imaged M31's disk over a $0.5~ \rm{deg}^2$ footprint with the HST, and is extensively described in
\cite{2012ApJS..200...18D}. The observations were divided into 23 bricks each consisting of 18 fields. The PHAT team provides photometry for over 117 million stars in six different HST filters \citep{2014ApJS..215....9W}, with supplementary data on each star's position, magnitude, magnitude error, chi values, signal-to-noise ratio, crowding, roundness and sharpness as determined from the PSF photometry package DOLPHOT \citep{2000PASP..112.1383D}. We downloaded the final catalog from the Mikulski Archive for Space Telescopes (MAST).\footnote{\url{https://archive.stsci.edu/prepds/phat/}}

We utilized photometry from the F814W and F110W filters, as the (F814W$-$F110W) color has been shown to be effective in identifying the JAGB stars in the HST photometric system \citep{2022ApJ...926..153M, lee2022}. We limited our photometry to the stars that satisfied the PHAT team's good star (GST) quality criterion in F814W and F110W, which was designed by the PHAT team to eliminate background galaxies and objects affected by blending, cosmic rays, or instrument artifacts. This criterion required a given star to have a signal-to-noise larger than 4, a sharpness squared value less than 0.15 and 0.20 in F110W and F814W, respectively, and a crowding parameter less than 1.30~mag and 2.25~mag in F110W and F814W, respectively.  These photometric quality cut combinations were deemed by the PHAT team to have produced the simultaneously deepest CMDs with the clearest features (e.g., upper MS, HeB sequence).

\subsection{Exclusion of Inner Metal-rich PHAT Bricks}\label{subsec:exclusion}

\cite{lee2022} showed that the JAGB method should be utilized in the outer disks of galaxies to minimize crowding and reddening of the JAGB stars.
In addition to avoiding these effects, in this section we explain why we were motivated to only utilize the outermost six bricks of the PHAT photometry due to contamination from metal-rich red giant branch (RGB) stars and low completeness in the inner disk/bulge.

Many of the RGB stars located in the inner bricks of the PHAT footprint are extremely metal-rich, causing them to be redder than metal-poor RGB stars \citep{2015ApJ...814....3D}. 
Furthermore, large photometric errors due to the high stellar density of the PHAT photometry also can cause the RGB stars to drift both bluer and redder. Ideally, the JAGB stars can be isolated through simple color cuts; however, doing so becomes increasingly difficult in the bulge and inner disk ($R<12$~kpc) of M31 where the RGB stars are as red or redder than the JAGB stars due to their high metallicity and photometric errors. 

Furthermore, the stellar density in the inner bricks is so high that the photometric bias increases and the completeness decreases. 
\cite{2015AJ....150..189G} created a 2D map of the RGB star completeness using the artificial star catalog provided by the PHAT team. An injected artificial star was defined as recovered if it passed certain photometric cuts and had an absolute magnitude change less than 0.75~mag. 
The completeness was then defined as the ratio of recovered artificial stars to injected artificial stars, where 100\% completeness meant all injected stars were recovered. \cite{2015AJ....150..189G} found the only bricks with more than 90\% completeness for the RGB stars were bricks 18-23 (see their Figure 6). 
Therefore, we chose to only use the photometry in bricks 18-23 to avoid the low-completeness bricks and metal-rich RGB stars. 

We also excluded brick 22 field 8, which showed strong contamination from Earth-limb glow in the IR exposures \citep{2014ApJS..215....9W}. This scattered light caused problems with the PSF fitting and background subtraction.

\subsection{Background Galaxy and Milky Way Foreground Star Contamination}
As described in Section 4 of \cite{2014ApJS..215....9W}, after the GST sharpness cuts, background galaxies are expected to number $<<1\%$ of the total stars in the PHAT catalog. They also found that Milky Way foreground stars occupy the color range $0.4<F110W-F160W<0.8$~mag in their simulations. This color range is much bluer than the JAGB stars, and thus Milky Way foreground stars do not contaminate the JAGB star population.

\subsection{Crowding and Artificial Star Tests}

We also quantified the completeness and potential photometric biases due to crowding in our photometry. \cite{2014ApJS..215....9W} computed artificial stars for the PHAT photometry. Because these tests are so computationally expensive (2000 CPU hours to run 50,000 stars for each field), artificial stars were only computed in six PHAT fields which covered the full range of RGB stellar densities. We refer the reader to \cite{2014ApJS..215....9W} for full details, but in short, artificial stars were injected into the HST images and then photometry was re-run to determine if a given artificial star was recovered. 
One of their six chosen fields, brick 21 field 15, fell within the six outermost PHAT bricks used in our study. 
The catalog for this field contained 49,998 artificial stars. Only 58 artificial stars lied within the JAGB color range of $2.1<(F814W-F110W)<2.5$~mag (see Section \ref{subsubsec:color} for an exploration of these color limits). We measured 100\% completeness for the 20 stars in the magnitude range $16<F110W<20$~mag (note that the JAGB stars are all brighter than $F110W=20$~mag, see Figure \ref{fig:CMDs}), meaning every injected star was recovered. We found 95\% completeness for the 38 stars fainter than $F110W>20$~mag.

\begin{figure}
\centering
\includegraphics[width=\columnwidth]{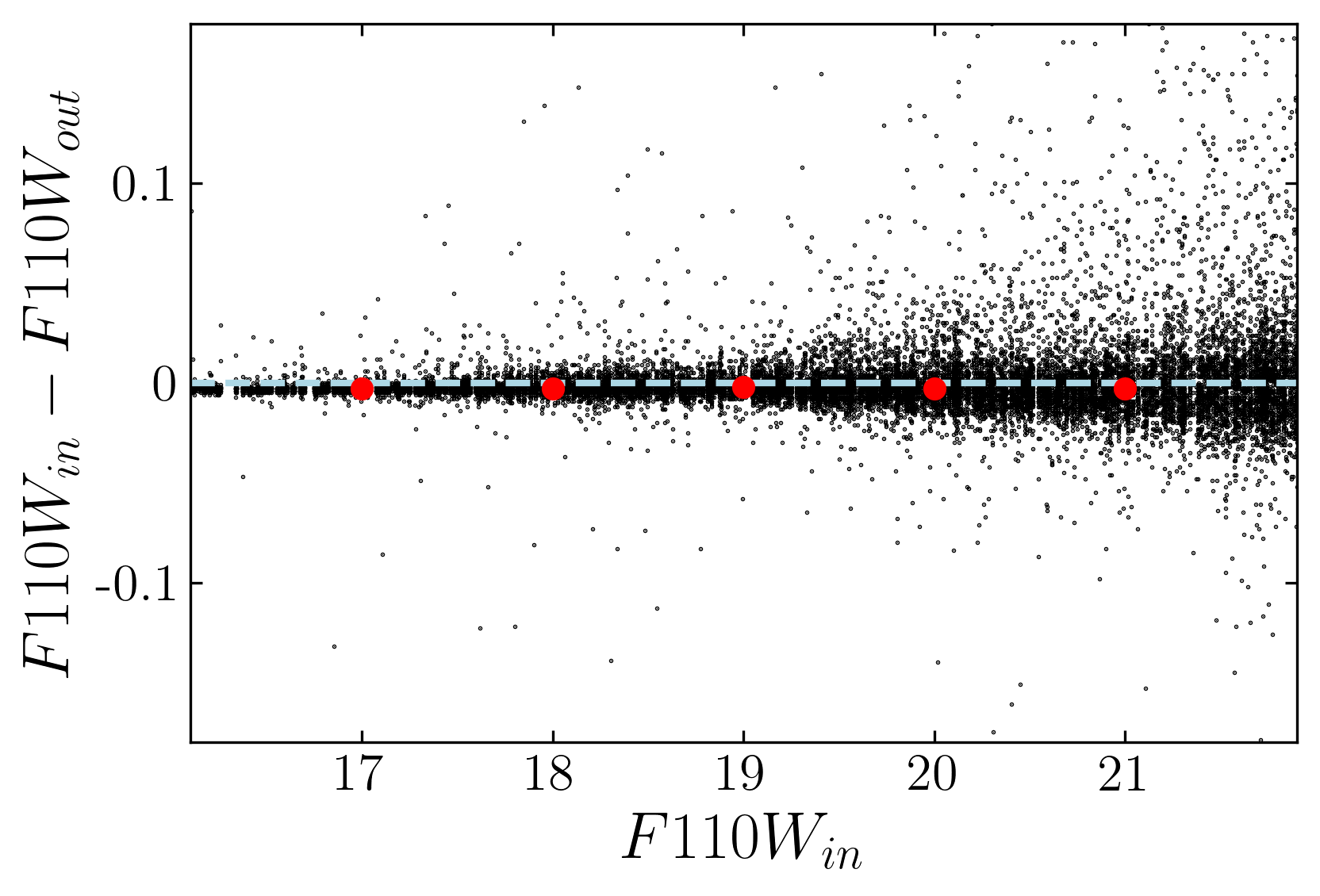}
\caption{The difference between the injected and recovered magnitudes for the entire catalog of artificial stars in brick 21 field 15 from \cite{2014ApJS..215....9W}. The measured median offsets in bins of 1~mag, shown as red points, were all less than 0.003~mag.}
\label{fig:offset}
\end{figure}

We also computed the median offset between the injected and recovered stars for the entire sample in Brick 21 field 15. The offsets for all the stars and the median offset in bins of 1~magnitude are shown in Figure \ref{fig:offset} as black and red points, respectively. Our measured offsets agreed exactly with those measured in \cite{2014ApJS..215....9W} (see their Table 7 and Figure 20). At the approximate level of the JAGB magnitude, the measured offset was $-0.002\pm0.0002$~mag.
This measured offset combined with the high completeness indicate crowding is a negligible source of bias in M31 for the JAGB stars, especially at the JAGB magnitude of $\sim18.8$~mag. 

\subsection{Ground-based J-band Comparison}
To check the accuracy of the distance modulus measured using HST WFC3/IR data in the F110W filter, we also measured a JAGB magnitude in the J band, which has been more conventionally used with the JAGB method. 
We established this J-band distance modulus as the fiducial distance because it was measured much farther out in M31's disk than the F110W distance moduli and therefore was significantly less affected by crowding and reddening effects.  We then were able to inspect the local variation of the measured F110W JAGB distances from the PHAT photometry compared to the measured ground-based distance.  

We utilized JK data from the Wide Field Camera (WFCam) taken from 2005 to 2008 on the 3.8-meter United Kingdom Infra-Red Telescope (UKIRT), kindly processed and then provided to us by Tongtian Ren and Yi Ren from their paper on red supergiants in M31 \citep{2021ApJ...907...18R}. This catalog originally had 1245930 sources. We kept only sources classified as \textit{stellar} in both J and K by the standard Cambridge Astronomy Survey Unit (CASU) pipeline, which uses the object's curve of growth to calculate a stellarness-of-profile statistic. 

Next, we calibrated the photometry using bright stars from the Two Micron All-Sky Survey (2MASS;  \citealt{2006AJ....131.1163S}). While WFCam photometry is nominally calibrated to 2MASS, we found small additional corrections were required for this catalog: $+0.032$~mag in the J band and $-0.005$~mag in the K band (2MASS $-$ WFCam). Similar photometric corrections were found by \cite{2020ApJ...889...44N} in their M31 WFCam JK photometry. We applied these corrections to our catalog.

We further cleaned the catalog using the photometric error as a function of magnitude in the J and K bands. We used the following constant+exponential functions, shown in Figure \ref{fig:error}, to clean the data, consistent with Chicago-Carnegie Hubble Program procedure (e.g., see the appendix in \citealt{2019ApJ...885..141B}):  

\begin{equation}
    \sigma_J < 0.02 + 0.003 \times e^{m_J - 16} 
\end{equation}

\begin{equation}
    \sigma_K < 0.02 + 0.003 \times e^{m_K - 15}. 
\end{equation}

\begin{figure}
\centering
\includegraphics[width=\columnwidth]{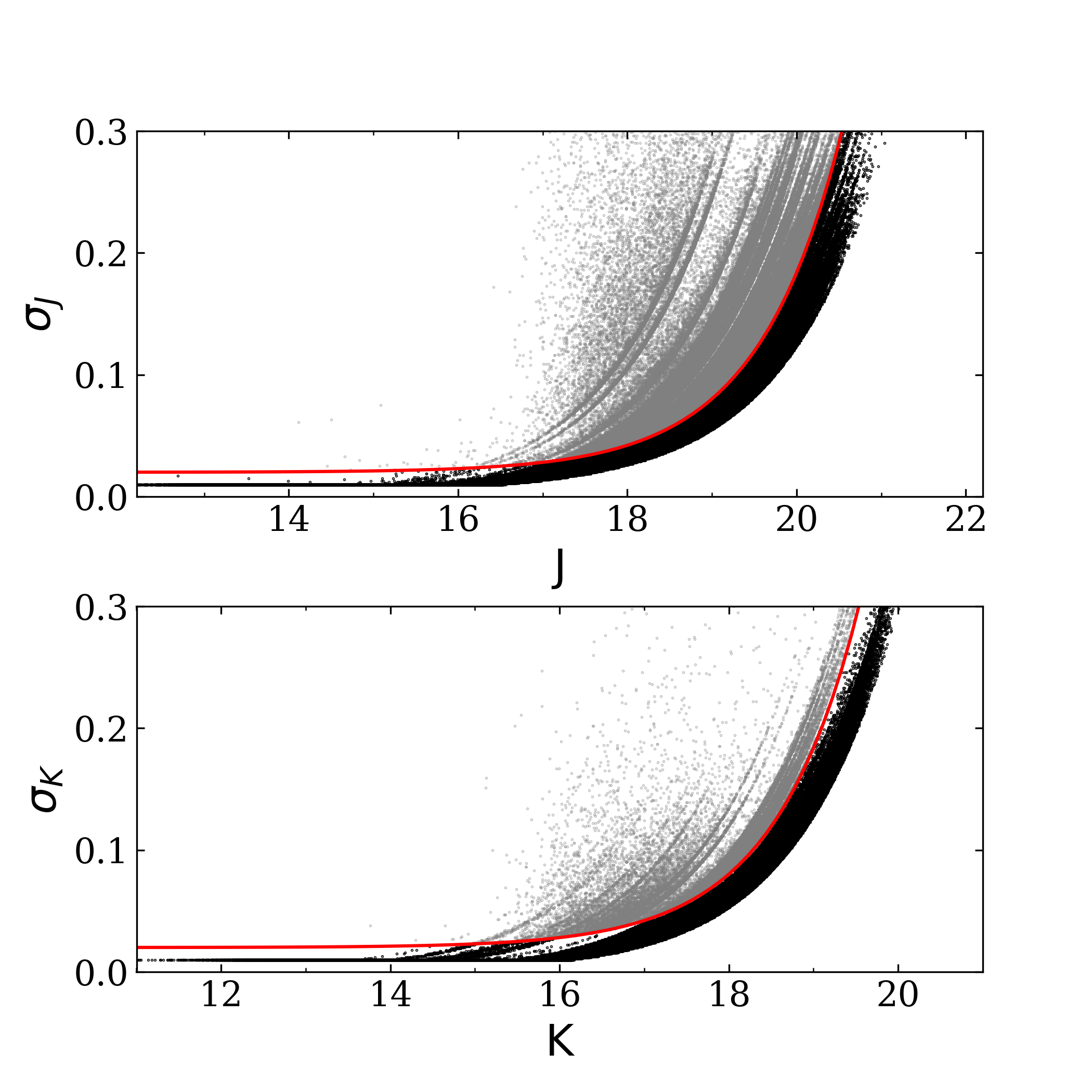}
\caption{Photometric uncertainty as a function of J- and K-band magnitudes. Quality cuts are shown in red, with the sources passing the given restriction in black and the sources not passing the restriction in grey.  } 
\label{fig:error}
\end{figure}

\begin{figure}
\centering
\includegraphics[width=\columnwidth]{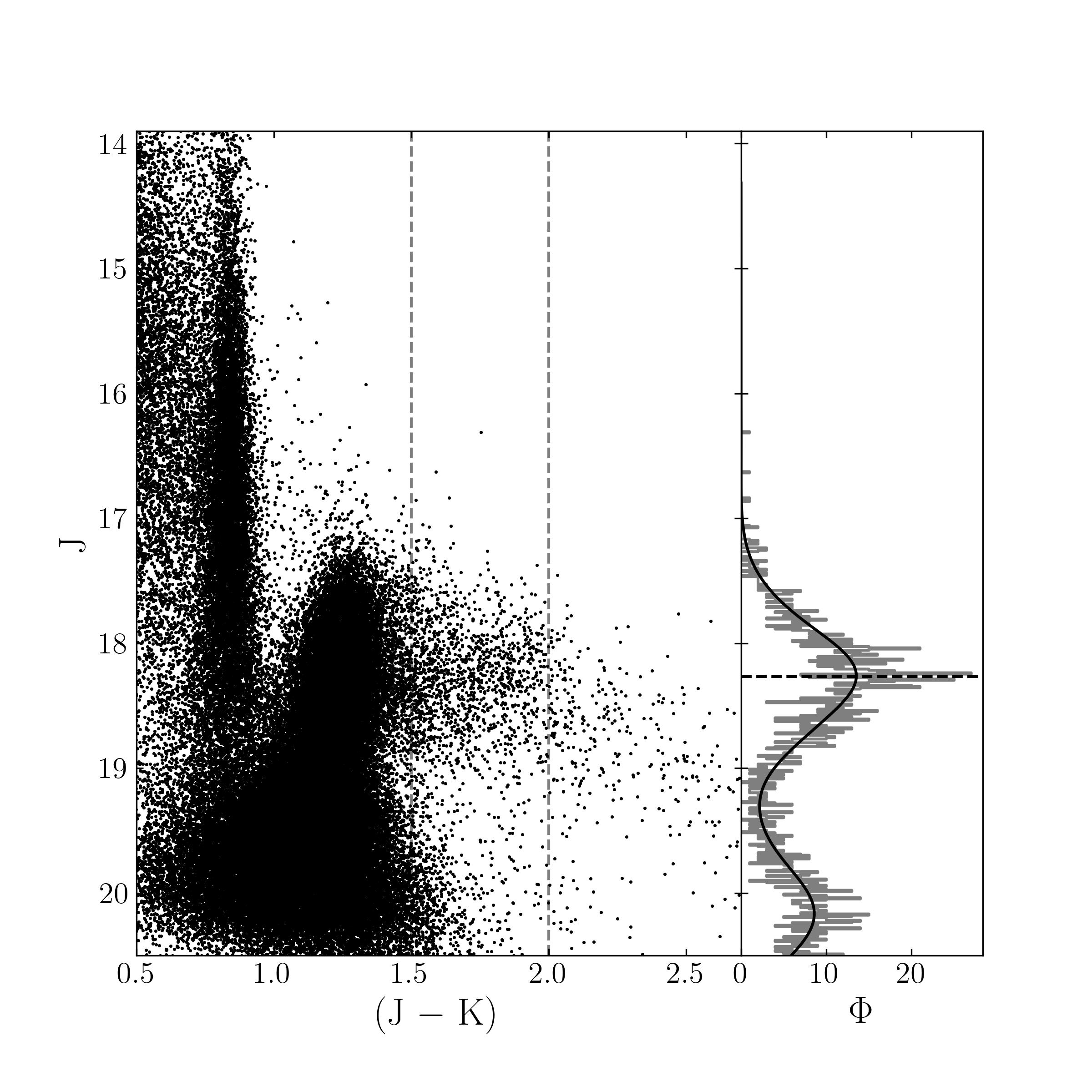}
\caption{UKIRT/WFCam color magnitude diagram for stars in M31's outer disk at $18<d<40$~kpc. The JAGB stars are located in the color range $1.5<(J-K)<2.0$~mag between the dotted lines. The smoothed LF in black is plotted over the binned LF in grey in the right panel, with the modal magnitude denoted by the dotted black line. The smoothing parameter was chosen to be $\sigma_s=0.38$~mag for visualization purposes; the choice of smoothing parameter varied the JAGB apparent magnitude by at most $\pm0.02$~mag.} 
\label{fig:JKcmd}
\end{figure}

\begin{figure*}
\centering
\includegraphics[width=.98\textwidth]{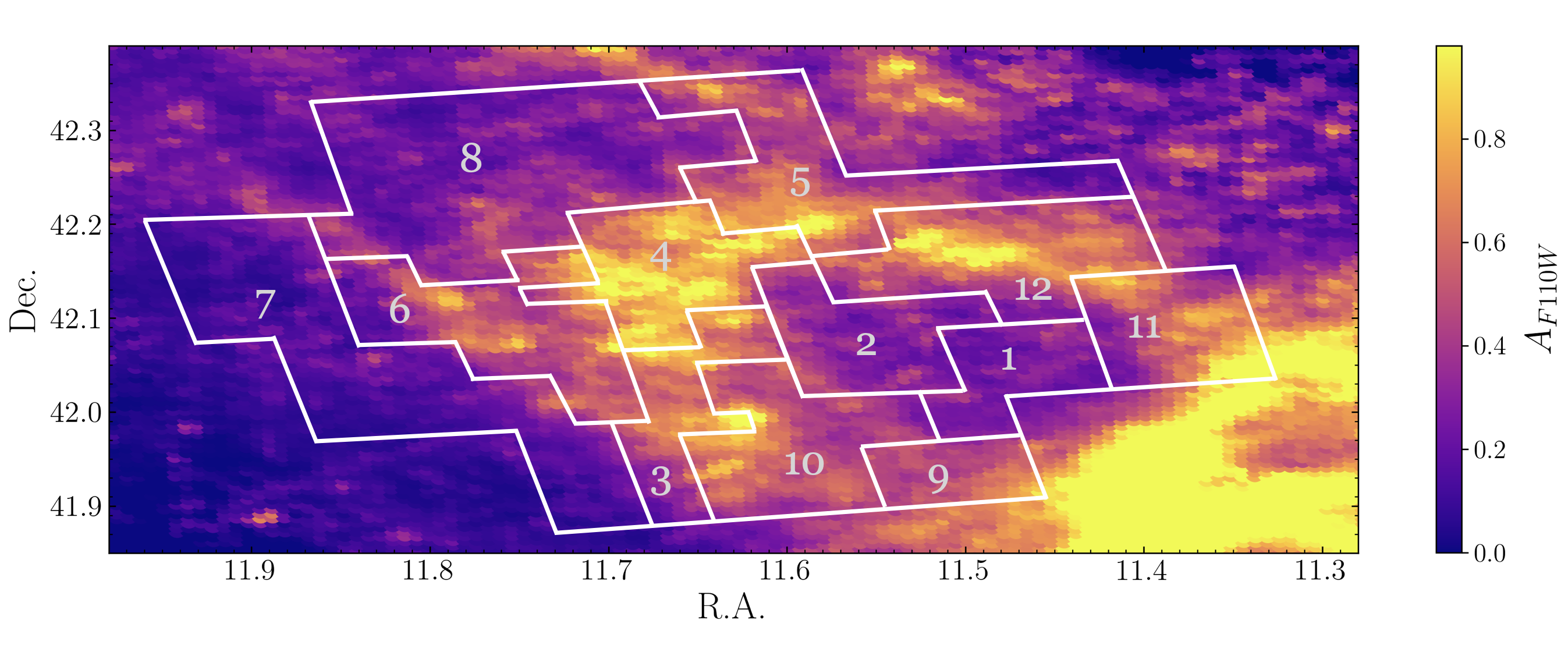}
\caption{Location of the 12 regions superimposed on the \cite{2014ApJ...780..172D} emission-based reddening map of M31, which is discussed in Section \ref{subsec:red}. The 12 regions cover approximately from 13 to 18 kpc of M31's northeast outer disk. The 12 regions were explicitly chosen to cover the most extreme range of reddening possible.}
\label{fig:dust} 
\end{figure*} 

\begin{deluxetable*}{cccccccc}\label{tab:data}
\tablecaption{Information on 12 Regions}
\tablehead{
\colhead{Region No.} & 
\colhead{No. JAGB stars} & 
\colhead{$m_{F110W}^{JAGB}$}  & 
\colhead{MAD ($\sigma$)} & 
\colhead{Skewness} & 
\colhead{$A_{F110W}$}  & 
\colhead{[M/H]} & 
\colhead{$N_{AGB}/N_{RGB}$}\\
\colhead{} & 
\colhead{} & 
\colhead{(mag)} & 
\colhead{(mag)} & 
\colhead{} &
\colhead{(mag)} & 
\colhead{(dex)} & 
\colhead{}
}
\startdata
1  & 265 & 18.70 & 0.23 & 0.10 & 0.27 & -0.25 & 0.239\\
2 & 295 & 18.79 & 0.27 & 0.14 & 0.36 &-0.23 & 0.234\\
3  & 252 & 18.90 & 0.36 & 0.20 &  0.60 &-0.20& 0.250\\
4&  283 & 18.87 & 0.31 & 0.21 &  0.74 &-0.20& 0.270\\
5& 270 & 19.07 & 0.32 & -0.01 &  0.59 &-0.19& 0.261\\
6&  225 & 18.79 & 0.24 & 0.13 & 0.48 & -0.22 & 0.221\\
7&  241 & 18.74 & 0.24& 0.16 &  0.17 &-0.26& 0.211\\
8&  283 &  18.84 & 0.25 & -0.02 &  0.32 &-0.25& 0.226\\
9&  249 & 18.79& 0.26 & 0.12 &  0.50 &-0.18& 0.249\\
10& 270 & 18.79 & 0.26 & 0.22 &  0.52 &-0.20& 0.240\\
11 & 291 & 18.93  & 0.35 & 0.16  & 0.58 &-0.18& 0.263\\
12 & 266& 18.83 & 0.33 & 0.12& 0.63 &-0.18& 0.268
\enddata
\end{deluxetable*}

The PHAT photometry extends only out to 18~kpc in M31's NE disk; the WFCam photometry extends significantly farther. We utilized a portion of the WFCam photometry between $18<d<40$~kpc, where reddening, blending, and crowding effects are minimal. This region of M31 is ideal for a JAGB measurement because the bulge population is completely negligible \citep{2003ApJ...588..311W} and there are still plentiful numbers of carbon stars; \cite{2005A&A...436...91D} found the number of carbon stars drops off sharply past 40~kpc. 
We also observed a population of reddened objects in the $40<d<100$~kpc stellar halo of M31 $\sim1$~mag fainter than the JAGB stars. These objects were similarly seen in the J vs. (J-K) color magnitude diagram from \cite{lee2022} in the outer regions of M33. These are likely extended sources like background galaxies (region L in \citealt{2000ApJ...542..804N}), and can be easily removed when we perform our own PSF photometry where the sharp and roundness parameters will be readily available from our own photometry pipelines. After making these photometric quality and spatial cuts, 92814 stars remained in the catalog.

In Figure \ref{fig:JKcmd}, we show the ground-based UKIRT/WFCam JK color-magnitude diagram. We selected the JAGB stars using color cuts of $1.5<(J-K)<2.0$~mag.\footnote{We plan to thoroughly investigate how the color selection of the JAGB stars changes in different environments in an exploration of the JAGB method in 13 nearby galaxies using observations obtained at the Magellan Telescope (Lee et al., in preparation).} As described in Section \ref{subsec:quantshape}, we then measured the apparent JAGB magnitude as the peak location of the smoothed JAGB star luminosity function. We measured the modal magnitude to be $m_{JAGB}=18.27\pm0.02$~(random)~mag. We also added in quadrature a random error accounting for the choice of smoothing parameter. Iterating through smoothing parameters between $\sigma_s$ of 0.10~mag and 0.50~mag in steps of 0.01~mag had at most a $\pm0.02$~mag effect on the modal magnitude. Thus, we adopted 0.02~mag as an additional statistical uncertainty on the final measurement. To measure the final distance modulus, we combined the apparent JAGB magnitude with the JAGB zeropoint from \cite{2022ApJ...938..125M} of $M_J=-6.20\pm0.01$ (stat) $\pm0.04$ (sys)~mag, which combines the geometric calibrations of the JAGB method from the LMC/SMC detached-eclipsing binaries \citep{2020ApJ...899...66M}, Milky Way Gaia DR3 parallaxes \citep{2021ApJ...923..157L}, and Milky Way open clusters \citep{2022ApJ...938..125M}. We also applied a correction for foreground extinction of $E(B-V)=0.062$~mag or $A_J=0.05$~mag \citep{2012ApJS..200...18D}. In conclusion, we measured a final distance modulus to M31 of $\mu_0 (J)=24.42\pm0.03$~stat~$\pm~0.04$ ~(sys)~mag. We will use this distance modulus as fiducial throughout the rest of this paper. This distance modulus agrees well with distances to M31 from the literature, which have remained largely uncontroversial in the last two decades, generally ranging between $\mu_0=24.40$~mag to $\mu_0=24.47$~mag \citep{2012ApJS..200...18D}. For example, recent distances from the literature have been measured via the Cepheids \citep{2021ApJ...920...84L}: $\mu_0 \rm{(Cepheids)}=24.41\pm0.03$~mag, the TRGB \citep{2005MNRAS.356..979M}: $\mu_0\rm{(TRGB)}=24.47\pm0.07$~mag, and detached-eclipsing binaries \citep{2005ApJ...635L..37R}: $\mu_0\rm{(DEB)}=24.44\pm0.12$~mag.

\section{JAGB Measurements in M31}\label{sec:jagbmeasure}

JAGB stars were selected using color cuts of $2.1<(F814W-F110W)<2.5$~mag. The blue color cut eliminates contamination from the the fainter metal-rich RGB stars and the red color cut excludes the red extreme carbon stars. We further explore the impact of these color limits on our final results in Section \ref{subsubsec:color}.
Next, we split the PHAT photometry into 12 regions explicitly chosen to span the a wide range of dust/reddening/age, to measure how the JAGB star luminosity function changes with stellar environment. 
In Figure \ref{fig:dust}, the boundaries of the 12 regions are shown overlaid on a reddening map of M31 (which is discussed in Section \ref{subsec:red}). Each region contained approximately 266 JAGB stars.

\subsection{Quantifying the shapes of JAGB star luminosity functions}\label{subsec:quantshape}

\begin{figure*}
\centering
\includegraphics[width=\textwidth]{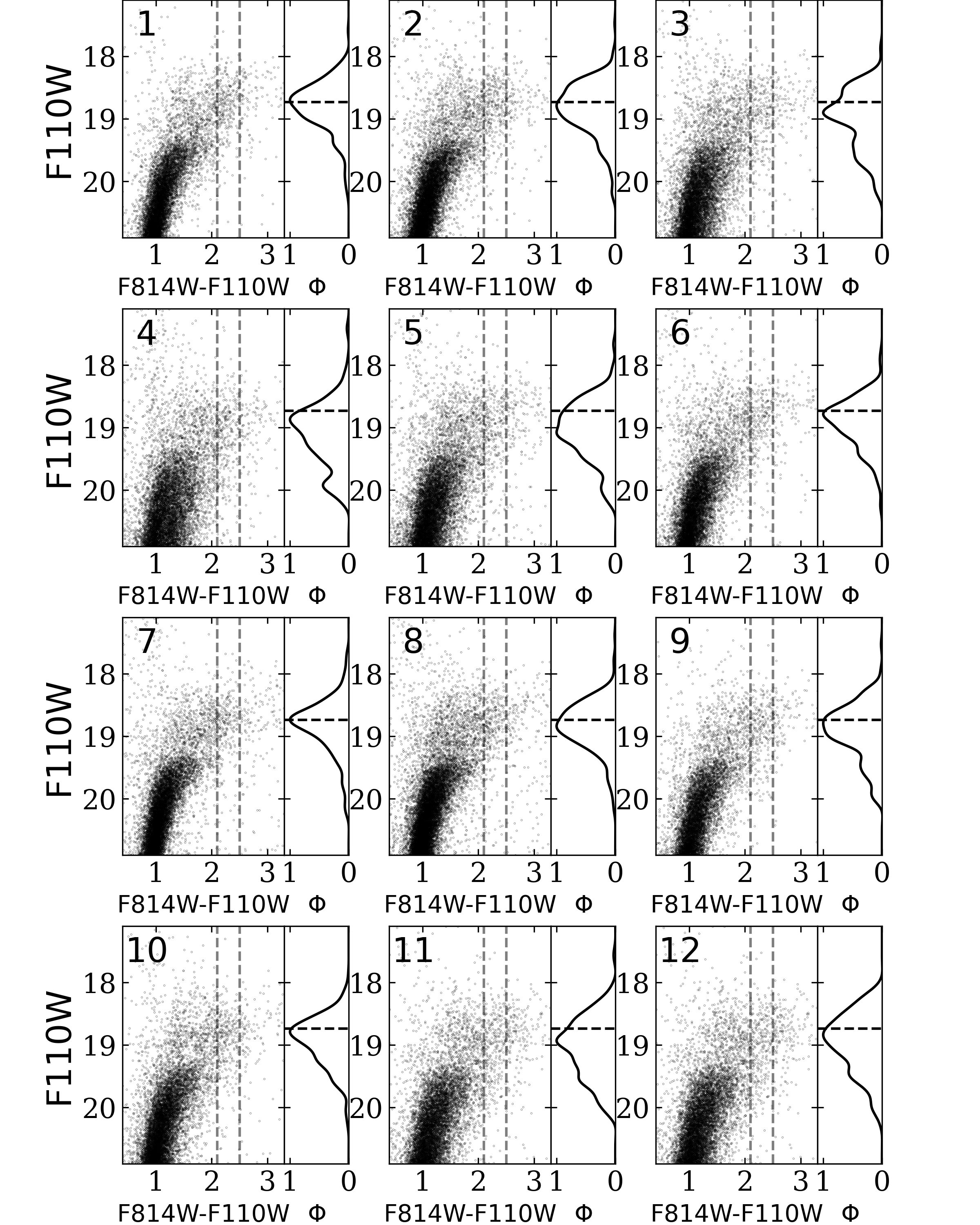}
\caption{F110W vs. (F814W$-$F110W) color-magnitude diagrams and smoothed luminosity functions for the 12 regions. The black dotted line marks the mode for the least reddened region 7. The locations of the 12 regions of are visually displayed in Figure \ref{fig:dust}.}
\label{fig:CMDs}
\end{figure*}

As in previous papers, we adopted the mode, or peak location of the JAGB star smoothed luminosity function as the apparent JAGB magnitude. The mode is less susceptible to outliers than the median or mean.

To calculate the mode of the luminosity function, we first binned the JAGB star F110W magnitudes using bins of 0.01~mag. To control for Poisson noise in the binned LF, we then smoothed it using the GLOESS (Gaussian-windowed, locally weighted scatterplot smoothing) non-parametric algorithm, a data-smoothing interpolating technique that is effective at suppressing false (noise-induced) edges and peaks in luminosity functions \citep{loess_ref, loader, 2004AJ....128.2239P}. The peak location of the smoothed luminosity function then marks the JAGB magnitude, $m_{F110W}^{JAGB}$. We chose a smoothing parameter of $\sigma_s$=0.20~mag for all 12 luminosity functions. The effect of varying smoothing parameters on the final result are explored further in Section \ref{subsubsec:smooth}. The final color-magnitude diagrams and smoothed luminosity functions for all 12 regions are shown in Figure \ref{fig:CMDs}.

To quantify the shape of the JAGB star luminosity function, we measured the mode, median absolute deviation $\sigma$, and skewness of each region's F110W luminosity function. The median absolute deviation is a robust measure of the spread of a population, and is more resistant to outliers in a dataset than the standard deviation. 
To measure the skew, we used the \textit{medcouple} statistic, a robust statistic that measures the skewness of a univariate distribution \citep{medcouple}. We measured the skewness of each JAGB star luminosity function to evaluate its asymmetry, as was done in \cite{2021MNRAS.501..933P, 2023MNRAS.522..195P} for their galaxy sample.
Standard errors for each parameter were calculated using 10,000 bootstrap realizations.

The measured skew, median absolute deviation, and mode $m_{F110W}^{JAGB}$ for each region can be found in Table \ref{tab:data}, along with each region's number of JAGB stars. We also list the average values of reddening, metallicity, and age in each region, which are all discussed in the following subsections.

\subsection{Reddening}\label{subsec:red}

In this section, we compare the mode, skew, and median absolute deviation to the average value of reddening in each region. For our reddening values, we used the reddening map from \citealt{2014ApJ...780..172D} (hereafter \citetalias{2014ApJ...780..172D}). This 2D map of the dust mass surface density, $\Sigma_{Md}$ ($M_{\odot}/\rm{kpc}^2$) in M31\footnote{Publicly available at \url{https://www.astro.princeton.edu/~draine/m31dust/m31dust.html}} is based on observations from the \textit{Spitzer Space Telescope} and \textit{Herschel Space Observatory}. \citetalias{2014ApJ...780..172D} used the physical dust model from \cite{2007ApJ...657..810D} and FIR and submillimeter emission observations to model $\Sigma_{Md}$, which can then be used to make a direct prediction of the line-of-sight extinction $A_V$ where $A_V/\Sigma_{Md} \cos i ~\rm{(mag/M_{\odot} pc^{-2})} = 7.394$ and cos~$i = 0.213$. 
To transform from $A_V$ to $A_{F110W}$, we adopted $A_{F110W}/A_V=0.337$ from \cite{2008PASP..120..583G}, assuming the \cite{1989ApJ...345..245C} extinction law and an $R_V=3.1$.\footnote{These scaling relations and extinction law ($R_V=3.1$) were also used by \cite{2015ApJ...814....3D} in their study of the extinction within M31. They found that using a significantly steeper attenuation law ($R_V=5$) increased $A_{F110W}/A_V$ by 0.05~mag (15\%), a change smaller than the typical precision (20\%) of the \citetalias{2014ApJ...780..172D} dust map. }
We utilized the map obtained from the MIPS160 camera ($18\arcsec$ pixels), which \citetalias{2014ApJ...780..172D} ``considers to be their best estimate for the dust mass.'' The $A_{F110W}$ map is shown in Figure \ref{fig:dust}.\footnote{We note that although \cite{2015ApJ...814....3D} found their independent CMD-derived extinction map of M31 predicted observed extinction about $\sim2.5$ times less than the \citetalias{2014ApJ...780..172D} map, their overall morphological agreement with the \citetalias{2014ApJ...780..172D} reddening map is excellent (see Figures 25-27 in \citealt{2015ApJ...814....3D}). Because we are only interested in relative changes in internal reddening throughout M31, our final results are unaffected by these absolute differences.
The observational tests presented in this study are differential and therefore differences between \textit{absolute} maps of reddening and metallicity are less significant as we are only interested in \textit{relative} changes over the disk of M31. However, when we used the \citetalias{2014ApJ...780..172D} map to correct the photometry for reddening, these absolute differences must be taken into consideration and are discussed in Section \ref{subsubsec:correct}.}
 
\begin{figure*}
\centering
\includegraphics[width=\textwidth]{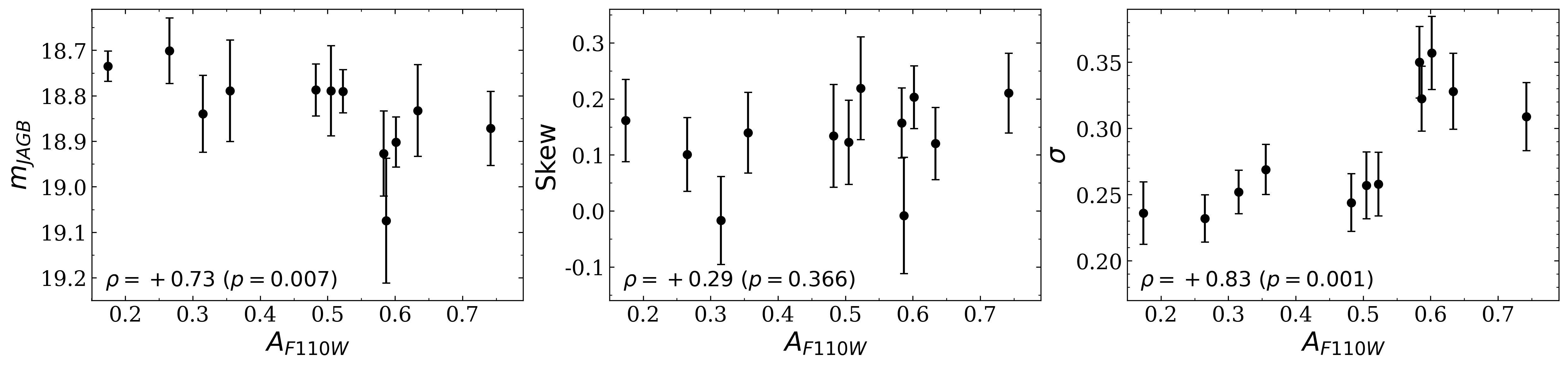}
\caption{Mode, skewness (medcouple), and median absolute deviation $\sigma$ parameters vs. line of sight reddening derived from \cite{2014ApJ...780..172D}. 
The Spearman's rank correlation coefficient and corresponding p-value are displayed in the bottom lefthand corner of each plot. Error bars were derived from 10,000 bootstrapped realizations. }
\label{fig:red}
\end{figure*}

To quantify the average reddening affecting the JAGB stars in each region, we first matched every JAGB star to the closest pixel in the \citetalias{2014ApJ...780..172D} map. We then averaged the reddening for all the stars in that region. 
In Figure \ref{fig:red}, we show the mode, skewness, and median absolute deviation parameters vs. the average reddening in each region. We also calculated the Pearson correlation coefficient for every relationship.

As expected, the mode was found to be fainter in higher reddening regions on average; we measured a significant correlation between the mode and average reddening ($\rho = +0.73$, p = 0.007).
That being said, the distance moduli in the lowest reddening regions agree well with the J-band distance modulus measured in M31's outermost disk from $18<d<40$~kpc where reddening is expected to be minimal ($\mu_J=24.47\pm0.05$~mag).
In the lowest-reddening regions, region 1 ($A_{F110W}=0.27$~mag) and 7 ($A_{F110W}=0.17$~mag), we measured $\mu_{1}=24.47\pm0.07$ (stat)~mag and $\mu_{7}=24.51\pm0.03$ (stat)~mag, assuming the JAGB zeropoint of $m_{F110W}=-5.77$~mag \citep{2022ApJ...926..153M}. On the other hand, in the highest-reddening regions ($A_{F110W}>0.6$~mag), the JAGB magnitude was fainter by $\sim0.2$~mag compared with the lowest-reddening regions.
\cite{lee2022} hypothesized internal reddening was the cause of significant differences between the mode in the outer and inner regions of M33. We now see clear empirical evidence that JAGB measurements are most accurate in the outermost low-reddening disks of galaxies.

The JAGB star luminosity functions on average had larger spread in higher reddening regions. We measured a significant correlation between the median absolute deviation $\sigma$ and $A_{F110W}$ ($\rho = +0.83$, p = 0.001).  This effect was likely a result of differential reddening. In a region with high internal reddening, some stars will be in front of the dust columns and other stars will be behind the dust. This will increase the spread in the measured apparent brightness of the stars. The correlation between skew and $A_{F110W}$ was measured to be insignificant ($\rho=+0.29$, p = 0.366).

\subsubsection{Reddening-Corrected Photometry}\label{subsubsec:correct}
\begin{figure*}
\centering
\includegraphics[width=\textwidth]{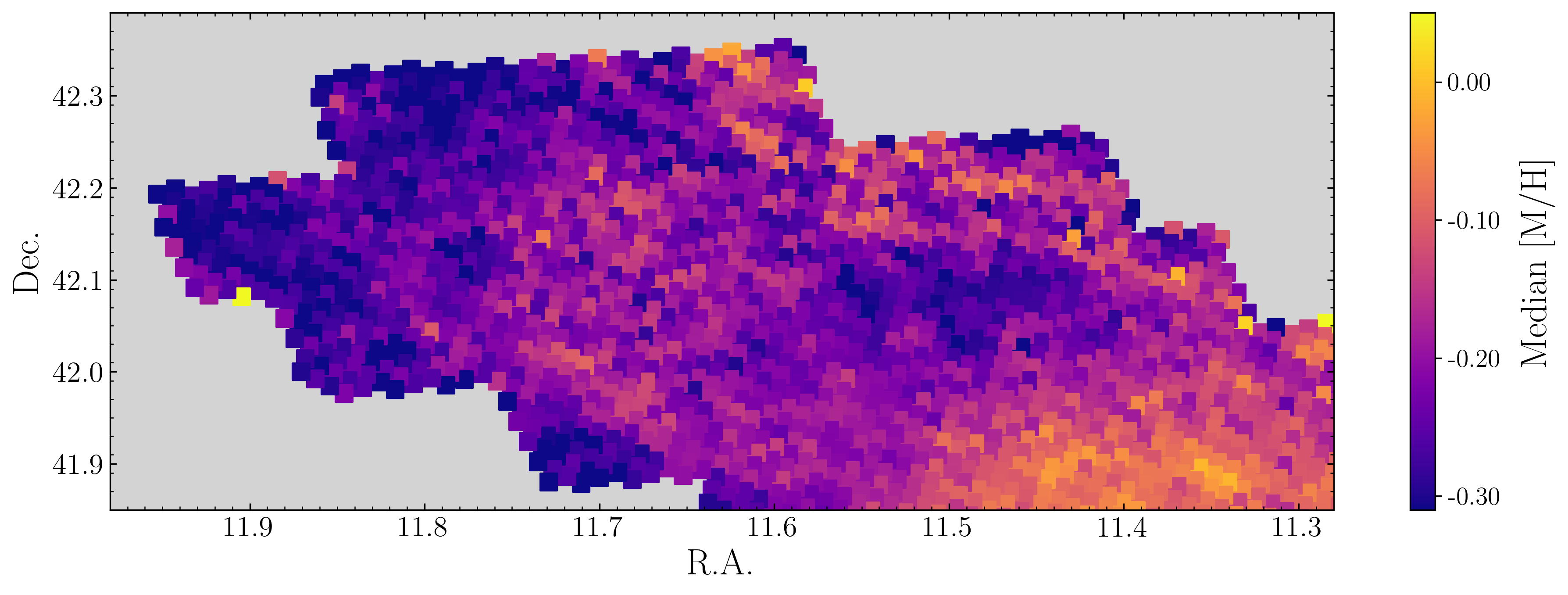}
\caption{Map of the \cite{2015AJ....150..189G} metallicities. }
\label{fig:metal}
\end{figure*}

In the following two sections, we quantify the dependence of the JAGB method on metallicity and age. However, first we disentangled any degenerate effects between reddening and age or metallicity. For example, reddening is clearly correlated with both metallicity and age, as shown through comparing Figure \ref{fig:dust} of the reddening in M31, Figure \ref{fig:metal} of the metallicity, and Figure \ref{fig:age_plot} of the age. Therefore, to separate covariant reddening effects with age and metallicity, we corrected the photometry for reddening in our metallicity and age analyses by subtracting $A_{F110W}$ from the F110W magnitude of the JAGB stars.

Using the \citetalias{2014ApJ...780..172D} emission-based reddening map to correct our photometry for line-of-sight extinction could be problematic. \cite{2015ApJ...814....3D} found their reddening-based map of M31 predicted less extinction than the \citetalias{2014ApJ...780..172D} map by a factor of 2.53, potentially due to an issue with the calibration of the dust emission models used to derive the \citetalias{2014ApJ...780..172D} map. Overcorrecting our photometry for extinction would cause our measured distance moduli to be too bright. Thus, we can determine whether the \citetalias{2014ApJ...780..172D} or \cite{2015ApJ...814....3D} reddening map is more accurate by comparing the F110W corrected distance moduli with our ground-based distance modulus which was significantly less affected by reddening. 

Our ground-based distance modulus corrected for foreground extinction was measured to be $\mu_0 (J)=24.42\pm0.05$~mag. If we used the \citetalias{2014ApJ...780..172D} map to correct our photometry, the measured reddening-corrected F110W distance moduli ranged from $\mu_0=24.05$~mag to $\mu_0 = 24.33$~mag. If we used the \citetalias{2014ApJ...780..172D} map now divided by the scale factor of 2.53 calculated by \cite{2015ApJ...814....3D}, the reddening-corrected distance moduli ranged from $\mu_0=24.29$~mag to $\mu_0 = 24.56$~mag, much more in agreement with our measured $\mu_0 (J)$. Thus, we find the conclusions of \cite{2015ApJ...814....3D} to be correct: the \citetalias{2014ApJ...780..172D} reddening map likely over predicts extinction by a factor of $\sim2.5$. In conclusion, in the following sections when we correct our photometry for reddening, we utilized \citetalias{2014ApJ...780..172D} map divided by a factor of 2.53 to correct our photometry.\footnote{The \cite{2015ApJ...814....3D} reddening map of M31 is not publicly available and therefore we were unable to use it directly for our analysis.} We show results for both the reddening-corrected and reddening-uncorrected photometry in Figure \ref{fig:metalplot} and Figure \ref{fig:age}, but only discuss results for the reddening-corrected photometry.

\subsection{Metallicity}\label{subsec:metalllicity_effect}

For the metallicity analysis, we used the stellar metallicity [M/H] map of M31's disk from 
\cite{2015AJ....150..189G} (hereafter \citetalias{2015AJ....150..189G})\footnote{Kindly provided to us by Dylan Gregersen and Anil Seth.}, which was calculated by fitting the red giant branch population with the Padova isochrones from \cite{2012MNRAS.427..127B}.  
While these metallicity estimates are for an older (RGB stars) population than the intermediate-aged population (AGB stars) relevant here, the \citetalias{2015AJ....150..189G} metallicity gradient was found to be consistent with gradients probing the metallicities of younger populations, indicating the M31 metallicity gradient is stable over long timescales \citep{2019ApJ...879..109B}.
For example, several studies on M31 have found that the \citetalias{2015AJ....150..189G} [M/H] metallicities are significantly anticorrelated with the ratio of C-type AGB stars to M-type AGB stars (C/M) in the NE disk of M31, a ratio that has long been established as a robust probe of AGB star metallicities \citep{2015ApJ...810...60H, 2019ApJ...879..109B}. 
This trend occurs because (1) AGB stars formed in higher metallicity regions have more free oxygen molecules in their atmospheres which preferentially binds with the carbon to form CO, thereby preventing the carbon enhancement of the star's atmosphere, and (2) the lower efficiency of the third-dredge in higher-metallicity environments lowers the amount of carbon transported to the surface \citep{2002PASA...19..515K}. 
The \citetalias{2015AJ....150..189G} metallacities have also been found to significantly correlate with oxygen abundances [O/H] in HII regions from \cite{2012MNRAS.427.1463Z}, which probe the metallicity of present-day star-forming regions and younger stellar populations \citep{2015ApJ...810...60H}.
Therefore, we found the [M/H] ratio to be a robust probe of the effect of \textit{relative} metallicity on the JAGB stars, due to its consistent agreement with metallicity probes of younger populations.

\begin{figure*}
\centering
\includegraphics[width=\textwidth]{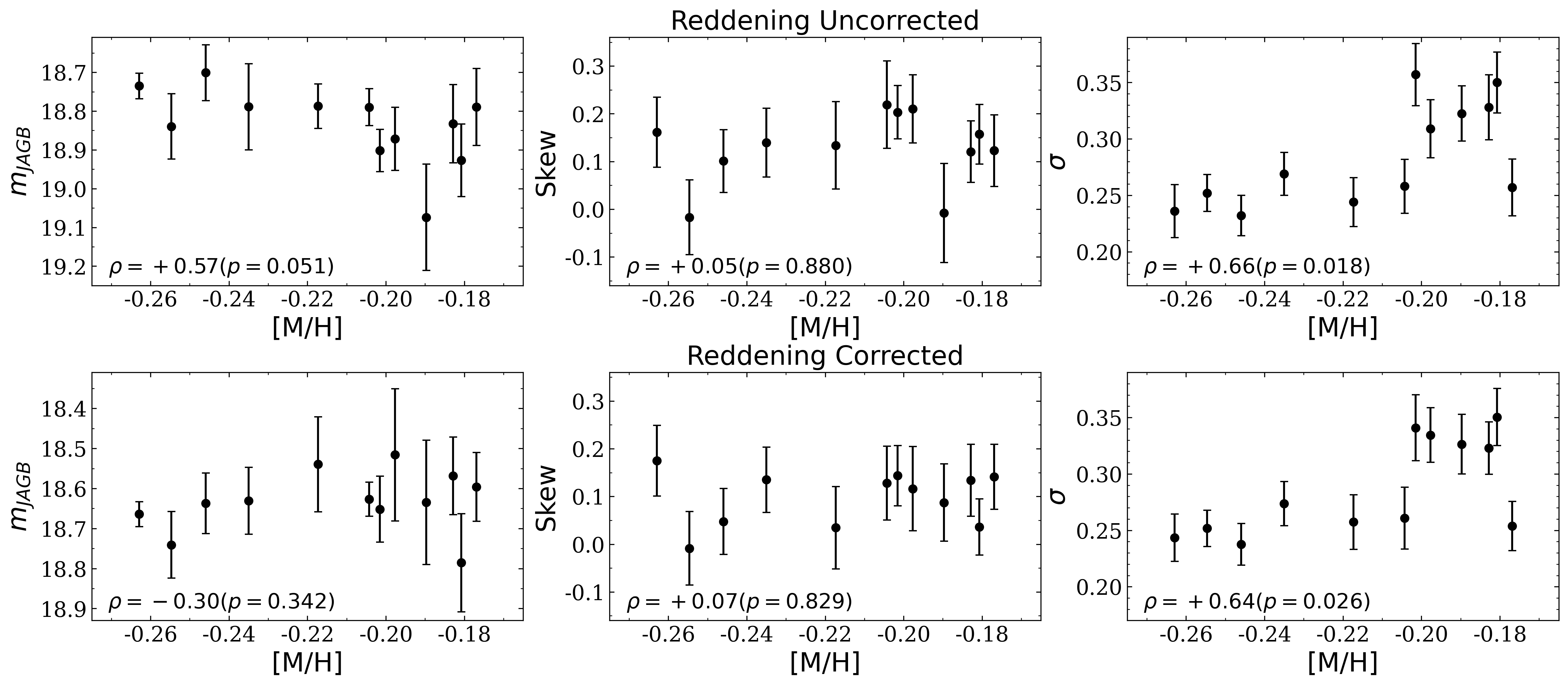}
\caption{Mode, skewness (medcouple), and median absolute deviation $\sigma$ parameters vs. metallicities derived from \citealt{2015AJ....150..189G}. 
The Spearman's rank correlation coefficient and corresponding p-value are displayed in the bottom lefthand corner of each plot. Error bars were derived from 10,000 bootstrapped realizations.  We show results for the reddening-uncorrected photometry on the top panels and the corrected photometry on the bottom panels.}
\label{fig:metalplot}
\end{figure*}

To quantify the average metallicity affecting the JAGB stars, we first matched every JAGB star to the closest pixel in the \citetalias{2015AJ....150..189G} map. We then averaged the [M/H] for all the stars in that region. A comparison is shown in Figure \ref{fig:metalplot} between the 
mode, skewness, and median absolute deviation of each region's luminosity function vs. the average metallicity of that region.

In the reddening-corrected case, we found no significant correlation between the mode and metallicity ($\rho=-0.30$, p = 0.342).
Therefore, our results show that any differences in metallicity had a negligible empirical effect on the observed mode of the JAGB star luminosity function for this metallicity range of $-0.18<[M/H]<-0.26$~dex.

We also note metallicity had a small impact on the spread ($\rho=+0.64$, p = 0.026). We hypothesize this is because M31's metal-rich RGB stars are so red that they contaminate the faint end of the JAGB star luminosity function. This would cause both an increase in the spread of the JAGB star luminosity function in higher-metallicity regions. We found no significant correlation between the metallicity and skew of the JAGB star LF ($\rho=+0.07$, p = 0.829).

\subsection{Age}

\begin{figure*}
\centering
\includegraphics[width=\textwidth]{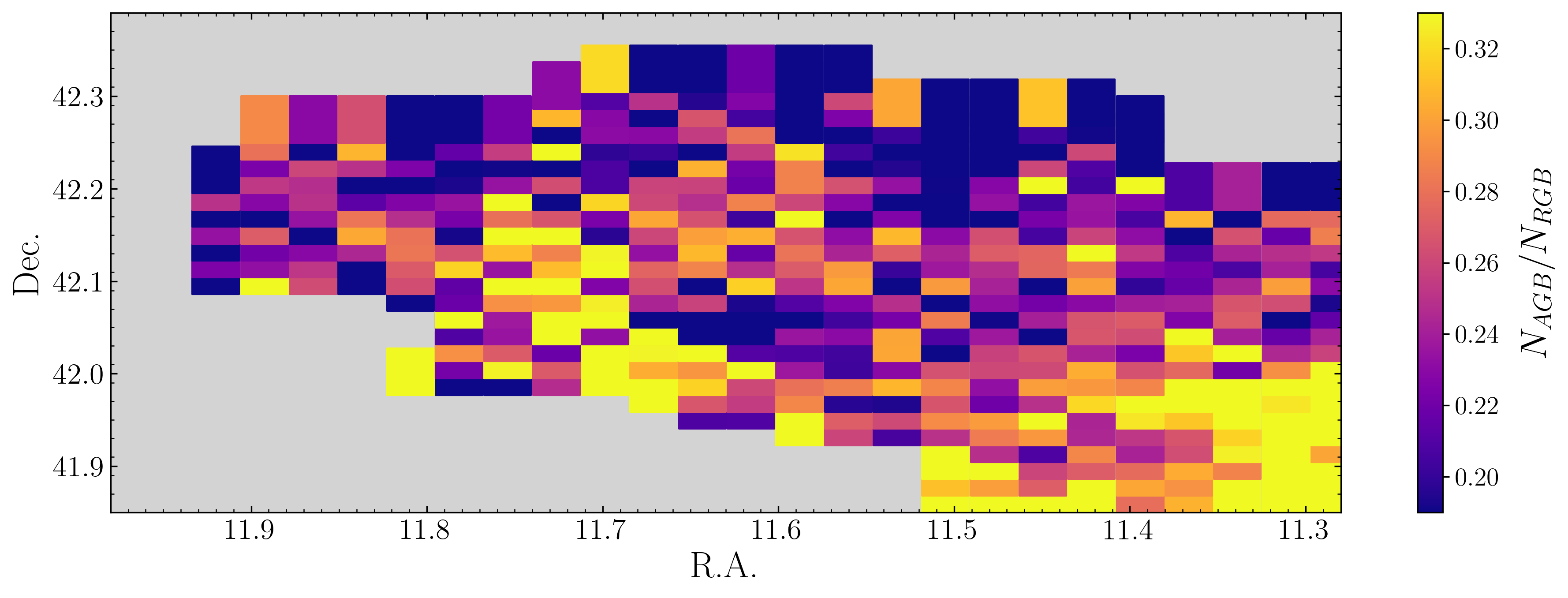}
\caption{Map of our average AGB star population age proxy $N_{AGB}/N_{RGB}$ in a given region. This ratio is expected to be lower in populations with a higher fraction of old stars (and thus an average older age) because TP-AGB stars have shorter lifetimes than RGB stars. For example, the darker purple areas represent older stellar populations with more relative numbers of RGB stars, where the most massive stars have already evolved through the AGB phase.
}
\label{fig:age_plot}
\end{figure*} 

To quantify the effect of the average AGB star age on the JAGB star luminosity function, we used the ratio of TP-AGB stars to RGB stars, $N_{AGB}/N_{RGB}$, which has been established as an approximate age proxy for the average TP-AGB star age in a given region \citep{2014ApJ...790...22R}. Because TP-AGB stars have smaller lifespans ($300$~Myrs to 1~Gyr) than RGB stars ($>4$~Gyrs), the average AGB star is younger than the average RGB star. 
A higher ratio of $N_{AGB}/N_{RGB}$ thus indicates a younger average population. 
Several studies have already used this ratio as a relative age proxy for the AGB stars in the PHAT photometry \citep{2015ApJ...810...60H, 2019ApJ...879..109B, 2022ApJS..259...41G}.

To photometrically classify the RGB and AGB stars, we followed the prescription of \cite{2022ApJS..259...41G}, who also used this $N_{AGB}/N_{RGB}$ ratio for the PHAT photometry to spatially map the average AGB star age. RGB stars were selected as having a color between $0.5<(F110W-F160W)<1.2$~mag and F110W magnitudes fainter than the TRGB and brighter than one magnitude below the TRGB (i.e. $19.28<F110W<20.28$~mag, \citealt{2019ApJ...879..109B}). 
While this stringent upper magnitude cut excludes fainter RGB stars, it also avoids completeness issues by restricting the magnitudes to well above the completeness limits in the PHAT photometry \citep{2022ApJS..259...41G}. Furthermore, including the fainter RGB stars in this ratio is unnecessary because we are only interested in comparing relative $N_{AGB}/N_{RGB}$ ratios between regions. AGB stars were selected also using the criteria from  \cite{2022ApJS..259...41G}:

\begin{enumerate}
    \item $F110W<19.28$~mag or $F160W<18.28$~mag (i.e. brighter than the TRGB)
    \item $F110W-F160W>0.88$~mag
    \item $F814W-F160W>2.4$~mag.
\end{enumerate}

\begin{figure*}
\centering
\includegraphics[width=\textwidth]{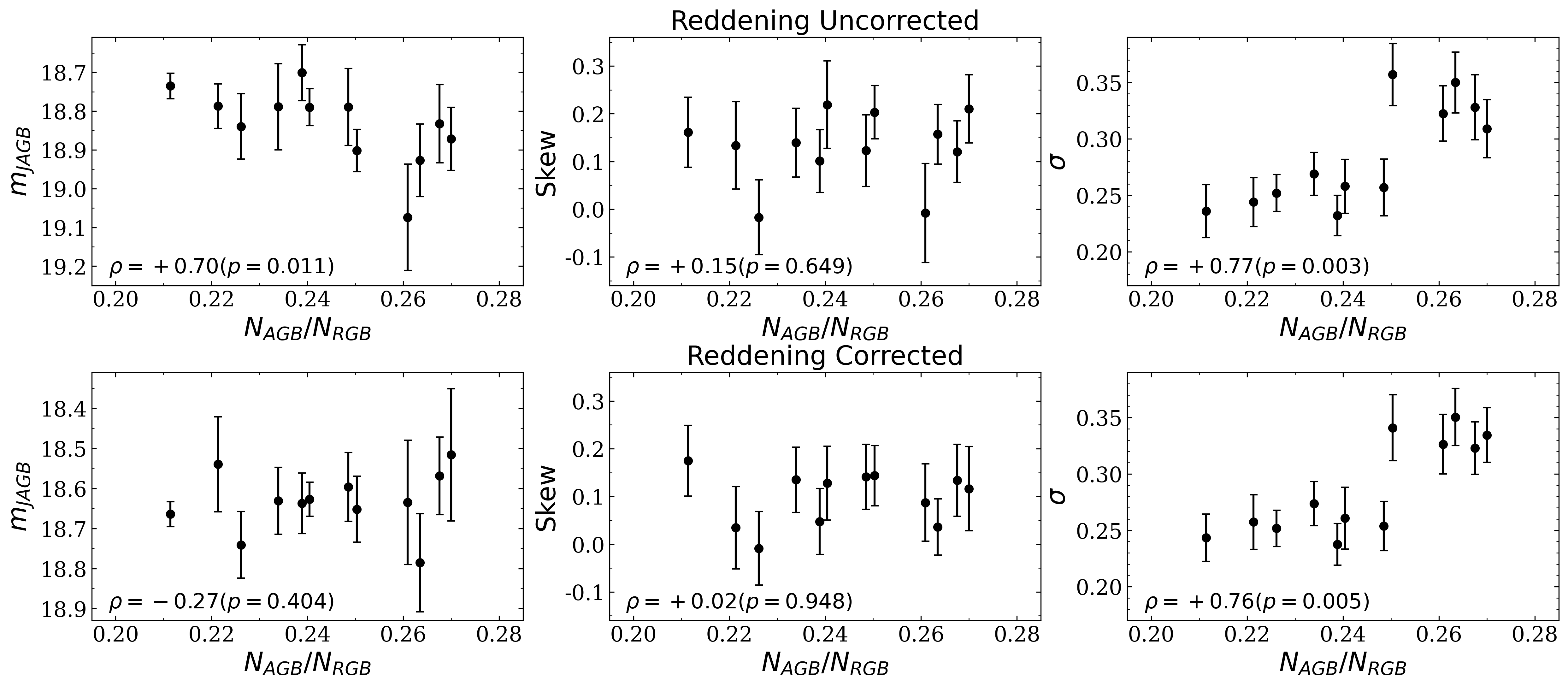}
\caption{Mode, skewness (medcouple), and median absolute deviation $\sigma$ parameters vs. $N_{AGB}/N_{RGB}$, a proxy for the average AGB star age in a region.
The Spearman's rank correlation coefficient and corresponding p-value are displayed in the bottom lefthand corner of each plot. Error bars were derived from 10,000 bootstrapped realizations.  We show results for the reddening-uncorrected photometry on the top panels and the corrected photometry on the bottom panels.}
\label{fig:age}
\end{figure*}

We calculated $N_{AGB}/N_{RGB}$ in each of the 12 regions using the PHAT photometry. For visual purposes, we display a spatially binned map of this ratio in Figure \ref{fig:age_plot}. In Figure \ref{fig:age}, we show the mode, skewness, and median absolute deviation parameters vs. the $N_{AGB}/N_{RGB}$ in each region. 

In the reddening-corrected case, the mode of the JAGB star luminosity function appears empirically to show little  dependence on the age of the population ($\rho=-0.27$, p = 0.404). These empirical results agree well with theoretical predictions that the average age of a stellar population should not significantly affect the JAGB star luminosity function. \cite{2008A&A...487..131C} inferred the best-fit mean age for the carbon stars in M33 via a comparison between the observed K-band luminosity function and the luminosity function derived from synthetic photometry from the \textsc{trilegal} population synthesis code. For the best-fit metallicity of Z = 0.0005, they found that differences in the mean age of the entire population from 2 to 10.6~Gyr had less than a 2\% difference  on the goodness-of-fit for comparing the theoretical with the observed carbon star luminosity function.

We also observed a significant correlation between the average age and scatter ($\rho=+0.76$, p = 0.005).
These correlations are likely because the reddening map and age map are spatially similar, as shown in a comparison of the reddening map in Figure \ref{fig:dust} and the age map in Figure \ref{fig:age_plot}. Therefore, the reasons discussed in Section \ref{subsec:red} are also relevant here. We found no significant correlation between the average age and skew ($\rho=+0.02$, p = 0.948).

\section{Ruling Out Potential Sources of Uncertainty}\label{sec:discussion}
In this section, we discuss tests for additional potential sources of uncertainty that may have affected our results.

\subsection{Contamination from Extrinsic Carbon Stars}
Interactions between binary stars are capable of producing \textit{extrinsic} carbon stars, i.e. carbon stars that are not produced by the third dredge-up (which are called intrinsic carbon stars). Instead, extrinsic carbon stars may form through the transfer of carbon-rich material from a companion (mass-losing AGB star transitioning to  a cool white dwarf). If binary populations are dependent on the environment (metallicity, age, etc.), it may affect the usefulness of carbon stars as distance indicators. However, these extrinsic carbon stars are much fainter and bluer than intrinsic carbon stars \citep{2016ApJ...828...15H}. In particular, \cite{2004MNRAS.350L...1I} both observationally and theoretically showed extrinsic carbon stars presented as a faint tail in the CSLF in the LMC and SMC, which peaks at a luminosity about $M_{bol}$= 1-2~mags fainter than the peak of the intrinsic CSLF. \footnote{In terms of relative numbers of extrinsic vs. intrinsic carbon stars: using I-band photometry of spectroscopically-identified carbon stars in the satellites and halo of M31, \cite{2016ApJ...828...15H} found that 42\% of their carbon star sample was brighter than the TRGB (i.e., expected to be intrinsic carbon stars), and 58\% was fainter (i.e., expected to be extrinsic carbon stars).}
Therefore, the existence of the much fainter extrinsic carbon stars is not expected to significantly affect the mode of the CSLF.

\subsection{Choice of smoothing parameter}\label{subsubsec:smooth}
The choice of smoothing parameter $\sigma_s$ when smoothing the luminosity function could potentially have an effect on the peak magnitude $m_{JAGB}$. 
However, we iterated through all smoothing parameters between [0.1, 0.2, 0.3, 0.4], and found that the slopes fit to the relations between $m_{JAGB}$ and $A_{F110W}$
remained correlated and statistically significant (i.e., $\rho>0.7$, $p<0.009$ for all cases) regardless of our choice of smoothing parameter. Similarly, using the reddening-corrected photometry we found the relations between $m_{JAGB}$ and both [M/H] and $N_{AGB}/N_{RGB}$ 
remained statistically insignificant (i.e., $|\rho|<0.3$, $p>0.3$ for all cases) regardless of our choice of smoothing parameter.
Therefore, we found that the choice of smoothing parameter had a negligible impact on our final conclusions. 

Other papers exploring the JAGB method have used different methods of measuring the JAGB apparent magnitude, each with their own advantages and disadvantages. The technique used in this paper requires a choice of smoothing parameter to measure the mode. \cite{2023MNRAS.522..195P} used a Lorentzian model fit by an un-binnned maximum likelihood estimator to measure the mode of the distribution. However, this method uses two different calibrators, the LMC and SMC, depending on the skewness of the galaxy's JAGB star luminosity function.\footnote{Because M31 is significantly more metal-rich than the Magellanic Clouds, it will first be necessary to extend this analysis to the more metal-poor M33 to illuminate whether using two different calibrations is necessary.} 
Extending this method to more metal-rich supernova host galaxies to measure $H_0$ could potentially introduce an unwanted systematic error. \cite{2021arXiv210502120Z} used a superposition of a Gaussian function and a quadratic function, obtaining the mean of the Gaussian via a non-linear least squares fitting method. This method also requires a user choice of binning scheme. Furthermore, the JAGB color cuts used in their work varied depending on the galaxy. Further work comparing the three methods in a sample of nearby galaxies, analyzed homogeneously and from the same telescope, would be useful.

\subsection{Color Selection of JAGB Stars}\label{subsubsec:color} 

The HST filter combination of (F814W$-$F110W) has been shown to be effective in isolating the JAGB stars and for accurately determining distances in the HST photometric system in two papers. First, \cite{2022ApJ...926..153M} preliminarily chose color limits of $1.5<(F814W-F110W)<2.5$~mag for the JAGB stars in a comparison of 20 JAGB and TRGB distances, finding a scatter of $\pm0.08$~mag between their distances. 
Second, based on the \textsc{colibri} theoretical isochrones \citep{2012MNRAS.427..127B}, \cite{lee2022} found $1.4<(F814W-F110W)<1.8$~mag were suitable color cuts for the JAGB stars in M33. Using these color cuts, \cite{lee2022} compared the measured JAGB magnitude in M33 using the HST (F814W$-$F110W) color with the more conventionally-used ground-based J-K color, finding 2\% agreement in their distances. 

However, contamination from RGB stars led us to move the M31 JAGB star color selection redder than what has been used in previous papers. The PHAT photometry has high stellar density, which causes large photometric errors as discussed in Section \ref{subsec:exclusion}. These photometric errors can cause the RGB to scatter redder. The RGB stars in M31 are also metal-rich, causing them to be reddened even further. 

We preliminarily chose the color selection of the JAGB stars to be $2.1<(F814W-F110W)<2.5$~mag.\footnote{We note that the (F814W$-$F110W) color of the JAGB stars is less stable than the (J-K) color throughout different galactic environments. The color of the JAGB stars in the HST color of (F814W$-$F110W) (approximately I - YJ) filters may potentially be more affected by environmental effects like metallicity, reddening, and crowding when compared with the ground-based J-K color. 
In a future paper (Lee, et al. in preparation), we plan to quantify how the color selection depends on host galaxy with a sample of 13 galaxies.} 
To test if the blue color cut significantly affected our final results, we shifted it within the range of [2.10, 2.15, 2.20, 2.25, 2.30]~mag and found that the slopes fit to the relations between $m_{JAGB}$ and $A_{F110W}$ remained correlated and statistically significant  (i.e., $\rho>0.70$, $p<0.01$ for all cases) regardless of our choice of color limits. Likewise, the relations between $m_{JAGB}$ and [M/H] and $N_{AGB}/N_{RGB}$ remained statistically insignificant ($p>0.2$ for all cases) through iterating through the different color cuts.

\section{Summary \& Future Work}\label{sec:conclusion}

\begin{enumerate}

    \item The \textit{mode} is significantly fainter in regions of high reddening. As a cross-check, we measured a ground-based J-band distance modulus of $\mu_J=24.47\pm0.05$~mag in the outermost disk of M31 ($18<d<40$)~kpc where internal reddening is expected to be minimal. This $\mu_J$ agrees well with the distance moduli measured in the lowest reddening regions 1 and 7 of $\mu_1=24.47\pm0.07$~mag and $\mu_7=24.51\pm0.03$~mag. The agreement demonstrates that JAGB method measurements are most accurate when performed in the outer disks of galaxies where reddening effects are minimized. 
    
    \item Reddening ($\rho=+0.83$, p = 0.001), metallicity ($\rho=+0.64$, p = 0.026), and age ($\rho=+0.76$, p = 0.005) all are also significantly correlated with the \textit{spread (median absolute deviation)} of the JAGB star luminosity function. Therefore, distances measured using JAGB stars in regions of higher-reddening, higher metallicity, and larger fractions of young AGB stars may be less precise. 
    
    \item  The mode of the JAGB star luminosity function appears empirically to show no dependence on the age or metallicity within the range of $-0.18<[M/H]<-0.26$~dex. We plan to extend this analysis to the more metal-poor M33 using the PHATTER data \citep{2021ApJS..253...53W} to increase the range of metallicities studied in this paper.

\end{enumerate}

In a future paper (Lee, et al. in preparation), we plan to carry out an additional investigation of the potential systematics on the JAGB method, instead using a sample of 13 galaxies from data uniformly obtained at the Magellan Telescope. This will extend this analysis to even a broader range of stellar environments, star formation histories, and metallicities.

\begin{acknowledgments}

I thank Wendy Freedman and Barry Madore for their valued feedback and suggestions.
I thank In Sung Jang for his never-ending patience in sharing his expertise on HST photometry.
I thank Meredith Durbin, Josh Frieman, Alex Ji, and Leslie Rogers for useful discussions that improved the analysis of this paper. I thank  Dylan Gregersen, Anil Seth, Tongtian Ren, and Yi Ren for generously providing me with their data of M31. I thank Bruce Draine on making his dust map of M31 publicly available. I thank the PHAT team on their incredible work and for making their exquisite photometry publicly available and easily accessible.  And finally, I thank the anonymous referee for their very constructive and helpful suggestions that improved this work.

This work is presented as (part of) a thesis to the Department of Astronomy and Astrophysics, The University of Chicago, in partial fulfillment of the requirements for the Ph.D. degree.

Financial support for this work was provided by NASA through grant No. HST-AR-17052 from the Space Telescope Science Institute, which is operated by AURA, Inc., under NASA contract NAS 5-26555. AJL was supported by the Future Investigators in NASA Earth and Space Science
and Technology (FINESST) award number 80NSSC22K1602 during the completion of
this work.
AJL was also partially supported through a NASA graduate fellowship grant awarded to the Illinois/NASA Space Grant Consortium.
AJL thanks the LSSTC Data Science Fellowship Program, which is funded by LSSTC, NSF Cybertraining Grant \#1829740, the Brinson Foundation, and the Moore Foundation; her participation in the program has benefited this work.

This research has made use of NASA's Astrophysics Data System Bibliographic Services. Some of the data presented in this paper were obtained from the Mikulski Archive for Space Telescopes (MAST) at the Space Telescope Science Institute. The PHAT data set can be accessed via \dataset[DOI: 10.17909/T91S30]{https://doi.org/10.17909/T91S30}.
\end{acknowledgments}

\facility{HST(ACS/WFC), HST(WFC3/IR), UKIRT (WFCam)}
\software{Astropy \citep{2013A&A...558A..33A, 2018AJ....156..123A}, NumPy \citep{2011CSE....13b..22V}, Matplotlib \citep{2007CSE.....9...90H}, Pandas \citep{pandas}, scipy \citep{2020NatMe..17..261V}, statsmodels \citep{statsmodels}.}

\end{document}